# New Perspective on the Reciprocity Theorem of Classical Electrodynamics

Masud Mansuripur[1,2)] and Din Ping Tsai[2)]

1) College of Optical Sciences, The University of Arizona, Tucson, Arizona 85721
2) Department of Physics, National Taiwan University, Taipei 106, Taiwan

**Abstract**. We provide a simple physical proof of the reciprocity theorem of classical electrodynamics in the general case of material media that contain linearly polarizable as well as linearly magnetizable substances. The excitation source is taken to be a point-dipole, either electric or magnetic, and the monitored field at the observation point can be electric or magnetic, regardless of the nature of the source dipole. The electric and magnetic susceptibility tensors of the material system may vary from point to point in space, but they cannot be functions of time. In the case of spatially non-dispersive media, the only other constraint on the local susceptibility tensors is that they be symmetric at each and every point. The proof is readily extended to media that exhibit spatial dispersion: For reciprocity to hold, the electric susceptibility tensor $\chi_{E\_mn}$ that relates the complex-valued magnitude of the electric dipole at location $r_m$ to the strength of the electric field at $r_n$ must be the transpose of $\chi_{E\_nm}$. Similarly, the necessary and sufficient condition for the magnetic susceptibility tensor is $\chi_{M\_mn} = \chi^T_{M\_nm}$.

**1. Introduction**. The principle of reciprocity in acoustic as well as electromagnetic (EM) systems was first enunciated by Lord Rayleigh [1]. Soon afterward, H. A. Lorentz and J. R. Carson extended the concept and provided sound physical and mathematical arguments that underlie the rigorous proof of the reciprocity theorem [2,3]. Over the years, the theorem has been embellished and extended to cover a broader range of possibilities, and to apply with fewer constraints [4-11]. The basic concept and its proof based on Maxwell's macroscopic equations are discussed in standard textbooks on electromagnetism [12,13]. For a recent review of reciprocity in optics, the reader is referred to the comprehensive article by Potton [14].

Roughly speaking, the idea of reciprocity can be stated as follows: In a linear, time-invariant electromagnetic system subject to certain restrictions, if the source of radiation is placed in region *A* and the resulting EM field is monitored in region *B*, then switching the locations of the source and the observer will result in the field monitored at *A* to be intimately related to that previously observed at *B*. Examples include:

i) The radiation pattern of an antenna is closely related to its reception pattern.

ii) Thin-film multilayer stacks of metals and dielectrics may have different reflectivities when illuminated from opposite directions (i.e., front illumination versus rear illumination), but they always exhibit precisely the same transmissivity at any given angle of incidence, irrespective of which side of the stack is illuminated [14,15].

iii) When a diffraction grating is illuminated at an arbitrary angle of incidence, several diffraction orders usually emerge, each propagating in a different direction and each having a specific diffraction efficiency. If now the direction of incidence is made to coincide with the path taken by one of the emergent orders, say, the one coming off at an angle $\theta_m$ and with a diffraction efficiency $\eta_m$, then one of the newly emerging diffraction orders will follow the previous path of incidence (in the reverse direction, of course), and will have the same efficiency $\eta_m$ [16].

This paper presents a simple yet general proof of the reciprocity theorem, which brings out the essential physics of the phenomenon and clarifies the need for the restrictions under which the theorem applies. The only mathematical fact needed in our proof is that the product of any number of matrices, say, $M_1, M_2, \ldots, M_n$, when transposed, will be equal to the product of the transposed matrices in reverse order, that is,

$$(M_1 M_2 M_3 \ldots M_{n-1} M_n)^T = M_n^T M_{n-1}^T \ldots M_3^T M_2^T M_1^T. \tag{1}$$

Generally speaking, reciprocity in electrodynamic systems applies when the material media that surround the source of radiation are linearly polarizable and/or magnetizable. The electric and magnetic susceptibility tensors of the material system can vary from point to point in space, but they must be time-independent. When the media are spatially non-dispersive, the only other constraint on the susceptibility tensors is the requirement of symmetry at each and every point. In other words, if the polarization density $P(r,\omega_o)$ and magnetization density $M(r,\omega_o)$ at a given point $r$ in space and at a fixed frequency $\omega_o$ are related to the local electric and magnetic fields, $E(r,\omega_o)$ and $H(r,\omega_o)$, through the following linear relations:

$$P(r,\omega_o) = \varepsilon_o \chi_E(r,\omega_o) E(r,\omega_o), \tag{2a}$$

$$M(r,\omega_o) = \mu_o \chi_M(r,\omega_o) H(r,\omega_o), \tag{2b}$$

then we must have $\chi(r,\omega_o) = \chi^T(r,\omega_o)$ for electric polarization (subscript $E$) as well as magnetization (subscript $M$). In the above equations, $\varepsilon_o$ and $\mu_o$ are the permittivity and permeability of free space, the system of units employed is MKSA, and both $\chi_E$ and $\chi_M$ are dimensionless entities. We shall impose no other restrictions on $\chi_E$ and $\chi_M$, allowing their components to be arbitrary complex-valued functions of $r$ and $\omega_o$.

The excitation source will be taken to be a stationary, monochromatic point-dipole, either electric or magnetic, located at $r_s$. The monitored field at the observation point $r_o$ will be either electric or magnetic, regardless of the nature of the source. When the source and the observed field are of the same type, i.e., both electric or both magnetic, one must use in the reverse path the same type of source and measure the same type of field as in the forward path. In contrast, when the source and the observed field are of different types, then, upon reversing the path, one must switch both the source type and the monitored field. For example, if the source at $r_s$ is an electric point-dipole, $p_o \exp(-i\omega_o t)$, while the observed field at $r_o$ is the $H$-field, then, in the reverse path, the source placed at $r_o$ must be a magnetic point-dipole, $m_o \exp(-i\omega_o t)$, and the field monitored at $r_s$ must be the $E$-field. (Note that the subscript "$o$" used in conjunction with the observation point $r_o$ is italicized. This should not be confused with the subscript "o" used with the frequency $\omega_o$ of the oscillations, with the amplitudes $p_o$ and $m_o$ of the dipoles, and with the EM field amplitudes $E_o$ and $H_o$.)

The paper is organized as follows. In Section 2 we describe the radiation field of an electric point-dipole in the surrounding free space, expressing the radiated EM fields in Cartesian coordinates for an arbitrary electric point-dipole $p_o \exp(-i\omega_o t)$ with components along the $x$-, $y$-, and $z$-axes. The corresponding formulas for the EM fields radiated by a magnetic point-dipole $m_o \exp(-i\omega_o t)$ are given in Section 3. In Section 4 we prove the reciprocity theorem in the simple case where the media surrounding the source are electrically polarizable, having a local, symmetric, time-independent electric susceptibility $\chi_E(r,\omega_o)$. The restriction to electrically polarizable media will be lifted in Section 5, where the media surrounding the source are allowed to have an electric as well as a magnetic susceptibility, $\chi_M(\mathbf{r},\omega_o)$. Section 6 generalizes the results to spatially-dispersive media, where both susceptibilities will be delocalized. In Sections 7 and 8 we remark on the differences between our approach to reciprocity and the conventional methods of proving the theorem. Section 9 summarizes the results of the paper and concludes with an observation regarding the Feld-Tai reciprocity lemma [8,9].

**2. Electromagnetic field radiated by an oscillating electric dipole**. With reference to Fig. 1, consider the electric point-dipole $\mathbf{p}(\mathbf{r},t) = p_{zo}\delta(\mathbf{r}-\mathbf{r}_s)\exp(-\mathrm{i}\omega_o t)\hat{\mathbf{z}}$, located at the fixed point $\mathbf{r}_s$ and oscillating along the $z$-axis with constant amplitude $p_{zo}$ and at fixed frequency $\omega_o$. At another point, say $\mathbf{r}_o$, the radiated fields given by an exact solution of Maxwell's equations are [17,18]:

$$\mathbf{E}(\mathbf{r}_o,t) = \mathbf{E}_o(\mathbf{r}_o - \mathbf{r}_s)\exp(-\mathrm{i}\omega_o t) = (p_{zo}/4\pi\varepsilon_o r) \tag{3a}$$
$$\times \{2\cos\theta\,[(1/r^2) - \mathrm{i}(\omega_o/cr)]\hat{\boldsymbol{\rho}} + \sin\theta\,[(1/r^2) - \mathrm{i}(\omega_o/cr) - (\omega_o/c)^2]\hat{\boldsymbol{\theta}}\}\exp[-\mathrm{i}\omega_o(t-r/c)],$$

$$\mathbf{H}(\mathbf{r}_o,t) = \mathbf{H}_o(\mathbf{r}_o - \mathbf{r}_s)\exp(-\mathrm{i}\omega_o t) = -(cp_{zo}/4\pi r)\{\sin\theta\,[(\omega_o/c)^2 + \mathrm{i}(\omega_o/cr)]\hat{\boldsymbol{\phi}}\}\exp[-\mathrm{i}\omega_o(t-r/c)]. \tag{3b}$$

In the above equations, $c = 1/\sqrt{\mu_o \varepsilon_o}$ is the speed of light in vacuum (with $\mu_o$ and $\varepsilon_o$ being the permeability and permittivity of free-space), $\mathbf{r} = \mathbf{r}_o - \mathbf{r}_s$ is the separation vector between the source dipole and the observation point, $r = |\mathbf{r}|$ is the length of $\mathbf{r}$, and $(\theta,\phi)$ are the polar and azimuthal coordinates of the observation point as seen from the location $\mathbf{r}_s$ of the dipole.

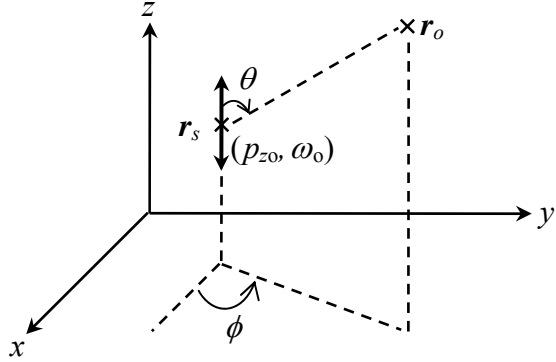

**Fig. 1**. An electric point-dipole located at $\mathbf{r}_s$ and oriented along the $z$-axis oscillates with frequency $\omega_o$ and (complex) amplitude $p_{zo}$. The electric and magnetic fields radiated by the dipole are observed at $\mathbf{r}_o$, whose position relative to $\mathbf{r}_s$ is given by $\mathbf{r} = \mathbf{r}_o - \mathbf{r}_s$. The spherical coordinates of the separation $\mathbf{r}$ between the source and the observer are $(r,\theta,\phi)$.

Defining

$$\Delta x = x_o - x_s, \tag{4a}$$

$$\Delta y = y_o - y_s, \tag{4b}$$

$$\Delta z = z_o - z_s, \tag{4c}$$

$$\cos\theta = (\Delta z)/r, \tag{4d}$$

$$\sin\theta = [(\Delta x)^2 + (\Delta y)^2]^{1/2}/r, \tag{4e}$$

$$\cos\phi = (\Delta x)/[(\Delta x)^2 + (\Delta y)^2]^{\frac{1}{2}}, \tag{4f}$$

$$\sin\phi = (\Delta y)/[(\Delta x)^2 + (\Delta y)^2]^{\frac{1}{2}}, \tag{4g}$$

we proceed to express the unit vectors $(\hat{\boldsymbol{\rho}}, \hat{\boldsymbol{\theta}}, \hat{\boldsymbol{\phi}})$ of the spherical coordinate system in terms of $(\hat{\boldsymbol{x}}, \hat{\boldsymbol{y}}, \hat{\boldsymbol{z}})$ of Cartesian coordinates, namely,

$$\hat{\boldsymbol{\rho}} = [(\Delta x)\hat{\boldsymbol{x}} + (\Delta y)\hat{\boldsymbol{y}} + (\Delta z)\hat{\boldsymbol{z}}]/r, \tag{5a}$$

$$\hat{\boldsymbol{\theta}} = \cos\theta\cos\phi\,\hat{\boldsymbol{x}} + \cos\theta\sin\phi\,\hat{\boldsymbol{y}} - \sin\theta\,\hat{\boldsymbol{z}}, \tag{5b}$$

$$\hat{\boldsymbol{\phi}} = -\sin\phi\,\hat{\boldsymbol{x}} + \cos\phi\,\hat{\boldsymbol{y}}. \tag{5c}$$

To simplify the notation further below, we define the following functions of $r$ and $\omega_o$:

$$f(r,\omega_o) = -(1/4\pi)[(1/r^3) - i(\omega_o/cr^2) - (\omega_o^2/c^2r)]\exp(i\omega_o r/c), \tag{6a}$$

$$g(r,\omega_o) = (1/4\pi)[(3/r^5) - i(3\omega_o/cr^4) - (\omega_o^2/c^2r^3)]\exp(i\omega_o r/c), \tag{6b}$$

$$h(r,\omega_o) = (1/4\pi)[(\omega_o^2/c^2r^2) + i(\omega_o/cr^3)]\exp(i\omega_o r/c). \tag{6c}$$

The EM fields of Eqs. (3), produced at $\boldsymbol{r}_o$ by the electric point-dipole located at $\boldsymbol{r}_s$, may now be written in Cartesian coordinates as

$$\boldsymbol{E}(\boldsymbol{r}_o,t) = \varepsilon_o^{-1} p_{zo} \{f(r,\omega_o)\hat{\boldsymbol{z}} + g(r,\omega_o)[(\Delta x \Delta z)\hat{\boldsymbol{x}} + (\Delta y \Delta z)\hat{\boldsymbol{y}} + (\Delta z)^2 \hat{\boldsymbol{z}}]\}\exp(-i\omega_o t), \tag{7a}$$

$$\boldsymbol{H}(\boldsymbol{r}_o,t) = c p_{zo} h(r,\omega_o)[(\Delta y)\hat{\boldsymbol{x}} - (\Delta x)\hat{\boldsymbol{y}}]\exp(-i\omega_o t). \tag{7b}$$

From the above expressions of the complex field amplitudes corresponding to a $z$-oriented dipole, one can readily determine the corresponding amplitudes when the point-dipole oscillates either along the $x$- or the $y$-axis. Consequently, the EM fields at $\boldsymbol{r}_o$, produced by the electric point-dipole $\boldsymbol{p}_o = p_{xo}\hat{\boldsymbol{x}} + p_{yo}\hat{\boldsymbol{y}} + p_{zo}\hat{\boldsymbol{z}}$ at $\boldsymbol{r}_s$, are given by the following matrix equations:

$$\begin{pmatrix} \varepsilon_o E_{xo} \\ \varepsilon_o E_{yo} \\ \varepsilon_o E_{zo} \end{pmatrix} = \begin{pmatrix} f(r,\omega_o) + g(r,\omega_o)(\Delta x)^2 & g(r,\omega_o)\Delta x \Delta y & g(r,\omega_o)\Delta x \Delta z \\ g(r,\omega_o)\Delta x \Delta y & f(r,\omega_o) + g(r,\omega_o)(\Delta y)^2 & g(r,\omega_o)\Delta y \Delta z \\ g(r,\omega_o)\Delta x \Delta z & g(r,\omega_o)\Delta y \Delta z & f(r,\omega_o) + g(r,\omega_o)(\Delta z)^2 \end{pmatrix} \begin{pmatrix} p_{xo} \\ p_{yo} \\ p_{zo} \end{pmatrix}, \tag{8a}$$

$$\begin{pmatrix} \mu_o H_{xo} \\ \mu_o H_{yo} \\ \mu_o H_{zo} \end{pmatrix} = Z_o \begin{pmatrix} 0 & -h(r,\omega_o)\Delta z & h(r,\omega_o)\Delta y \\ h(r,\omega_o)\Delta z & 0 & -h(r,\omega_o)\Delta x \\ -h(r,\omega_o)\Delta y & h(r,\omega_o)\Delta x & 0 \end{pmatrix} \begin{pmatrix} p_{xo} \\ p_{yo} \\ p_{zo} \end{pmatrix}. \tag{8b}$$

In Eq. (8b), $Z_o = \sqrt{\mu_o/\varepsilon_o} \approx 377\,\Omega$ is the impedance of the free space. Denoting the $3\times3$ coefficient matrices in Eqs. (8a) and (8b) by $U_{os}(\boldsymbol{r},\omega_o)$ and $V_{os}(\boldsymbol{r},\omega_o)$, respectively, the compact version of Eqs. (8) will be

$$\varepsilon_o \boldsymbol{E}_o(\boldsymbol{r}_o) = U_{os}(\boldsymbol{r},\omega_o)\boldsymbol{p}_o, \tag{9a}$$

$$\mu_o \boldsymbol{H}_o(\boldsymbol{r}_o) = Z_o V_{os}(\boldsymbol{r},\omega_o)\boldsymbol{p}_o. \tag{9b}$$

These *U* and *V* matrices have an important property that is of crucial significance for the reciprocity theorem. If the locations of the source and the observer are switched, that is, if the dipole is moved to $r_o$ and the observer placed at $r_s$, then $\Delta x$, $\Delta y$, and $\Delta z$ will change sign. As can be seen in Eq.(8a), since the coefficient matrix $U_{os}(r,\omega_o)$ contains products of *pairs* of $\Delta x$, $\Delta y$, and $\Delta z$, it will not change at all when the propagation direction is reversed. Furthermore, since $U_{os}(r,\omega_o)$ is symmetric, its transpose will be the same matrix. In contrast, the coefficient matrix $V_{os}(r,\omega_o)$ of Eq.(8b) contains elements that are proportional to $\Delta x$, $\Delta y$, or $\Delta z$; as such, the matrix will change sign upon reversing the propagation direction. However, $V_{os}(r,\omega_o)$ is antisymmetric, which means that its sign will change again when transposed. Both *U* and *V* thus have the essential property that when the propagation direction is reversed *and* the matrix is transposed, each matrix will remain intact. This feature of the coefficient matrices in Eqs.(9a) and (9b) is the fundamental physics behind the reciprocal properties of a wide class of linear, time-invariant electromagnetic systems.

**3. Electromagnetic field radiated by an oscillating magnetic dipole**. A magnetic point-dipole is the dual of an electric point-dipole, radiating an EM field similar to that given by Eqs.(8), albeit with the roles of *E* and *H* switched. To retain symmetry between the two types of dipole radiator, we define the magnetic moment *m* of a small loop of area *A* and current *I* such that its magnitude will be $m = \mu_o I A$. [This definition of the magnetic dipole is consistent with the *B*-field of Maxwell's equations being written as $B = \mu_o H + M$, as opposed to $B = \mu_o(H + M)$, which corresponds to the definition of the magnetic dipole moment as $m = IA$.]

For a magnetic point-dipole $m_o = m_{xo}\hat{x} + m_{yo}\hat{y} + m_{zo}\hat{z}$ located at $r_s$, the EM fields at $r_o = r_s + r$, obtained from an exact solution of Maxwell's equations are

$$\mu_o H_o(r_o) = U_{os}(r,\omega_o) m_o, \tag{10a}$$

$$\varepsilon_o E_o(r_o) = -Z_o^{-1} V_{os}(r,\omega_o) m_o. \tag{10b}$$

Aside from the role-reversal of $\varepsilon_o E_o$ and $\mu_o H_o$, the main difference between the fields in the above equations and those in Eqs.(9) is the change of the constant coefficient $Z_o$ of Eq.(9b) to $-Z_o^{-1}$ in Eq.(10b).

**4. Reciprocity in a system containing electrically-polarizable media**. Let the electric point-dipole $p(r_s,t) = p_o\exp(-i\omega_o t)$ be located at $r_s$ in the system depicted in Fig.2. This will be the only source dipole in the system, whose radiation will excite the electric dipoles of the surrounding media. The *E*-field arriving at the observation point $r_o$ is the superposition of the *E*-fields from all the dipoles in the system. The first dipole, $p_o$, being the source, has a fixed amplitude, of course, but the dipoles located at $r_1, r_2, \ldots, r_N$ will have their amplitudes determined by the local *E*-field. The dipoles at $r_1, r_2, \ldots, r_N$ represent the entirety of the linear media that surround the source and the observation point.

Taking a straight path from $r_s$ to $r_o$, the *E*-field that reaches directly from the source to the observer will be given by $\varepsilon_o E_o = U_{os} p_o$, where $U_{os}$ is the propagation matrix from $r_s$ to $r_o$. Next, take a different path from $r_s$ to $r_o$, this time visiting one or more dipoles of the surrounding media. Figure 2 shows a particular path that goes through $r_1, r_2, r_3$, and $r_4$, before arriving at $r_o$. Along this path, $p_o$ excites the dipole at $r_1$, which excites that at $r_2$, which excites that at $r_3$, which excites that at $r_4$, whose field then reaches the observer at $r_o$. The *E*-field that reaches $r_o$ from this particular path will be given by

$$\varepsilon_o E_o(\bm{r}_o) = U_{o4}\chi_{E\_4}U_{43}\chi_{E\_3}U_{32}\chi_{E\_2}U_{21}\chi_{E\_1}U_{1s}\bm{p}_o. \tag{11}$$

In the above equation, $\chi_{E\_n}$ is the electric susceptibility tensor associated with the dipole at $\bm{r}_n$. We shall assume that all such tensors are symmetric, but impose no other restrictions on them. For instance, the components of $\chi_{E\_n}$ may be real- or complex-valued, corresponding, respectively, to transparent or absorbing media.

Proceeding in similar fashion, we can compute the *E*-field at $\bm{r}_o$ produced by each and every dipole in the system, excited via all possible paths through the surrounding media. The total field at $\bm{r}_o$ will then be obtained by adding up all the fields produced via all possible paths from $\bm{r}_s$ to $\bm{r}_o$. Note that a given path, after leaving the source at $\bm{r}_s$ and before arriving at the observation point, may visit a given dipole (associated with the material media) any number of times, or it may not visit that particular dipole at all. The only dipole that must appear once and only once in the beginning of each and every path is the source dipole, because it is the source that initiates all other excitations, while its own excitation is not influenced in any way by radiation from the surrounding dipoles.

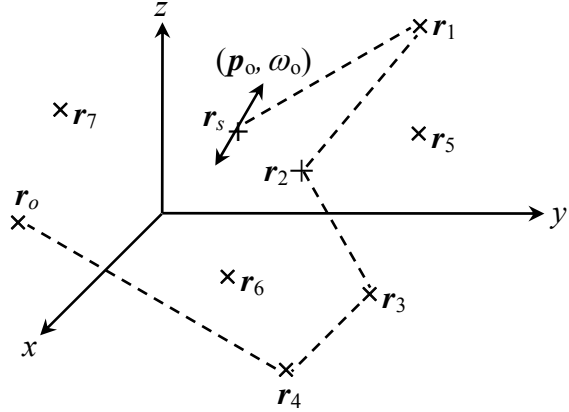

**Fig. 2**. An electric point-dipole located at $\bm{r}_s$ oscillates with frequency $\omega_o$ and complex amplitude $\bm{p}_o = p_{xo}\hat{\bm{x}} + p_{yo}\hat{\bm{y}} + p_{zo}\hat{\bm{z}}$. The system contains *N* other fixed dipoles, located at $\bm{r}_1, \bm{r}_2, \ldots, \bm{r}_N$, each of which responds linearly to the local EM field. The observation point $\bm{r}_o$ is where the electric field $\bm{E}(\bm{r}_o,t)$ is monitored. If the dipole attached to $\bm{r}_n$ is an electric dipole, it must respond to the local *E*-field with an electric susceptibility $\varepsilon_o\chi_{E\_n}(\omega)$ that is a symmetric tensor. Similarly, if the dipole attached to $\bm{r}_m$ is magnetic, it must respond to the local *H*-field with a magnetic susceptibility tensor $\mu_o\chi_{M\_m}(\omega)$ that is also symmetric. There is no specific relation between the susceptibility tensors associated with different dipoles of the system. The simplest path between the source and the observer is the direct path from $\bm{r}_s$ to $\bm{r}_o$. However, any path that starts at $\bm{r}_s$ and visits one or more of the dipoles before arriving at $\bm{r}_o$ is also legitimate. There is no limit as to how many times a particular dipole can be visited, nor are there any restrictions on the sequence of media dipoles that are included in a given path. The partial *E*-fields produced at $\bm{r}_o$ by the last dipole in each path must all be added together to produce the total *E*-field at the observation point.

As an example, consider a simple system consisting of a source dipole $\bm{p}_o$ located at $\bm{r}_s$, two electric dipoles with susceptibility tensors $\chi_1$ and $\chi_2$, located at $\bm{r}_1$ and $\bm{r}_2$, and an observation point $\bm{r}_o$, where the *E*-field is monitored. The first few terms of the infinite series that converges to the observed *E*-field are

$$\varepsilon_o E_o(\boldsymbol{r}_o) = U_{os}p_o + U_{o1}\chi_1 U_{1s}p_o + U_{o2}\chi_2 U_{2s}p_o + U_{o2}\chi_2 U_{21}\chi_1 U_{1s}p_o + U_{o1}\chi_1 U_{12}\chi_2 U_{2s}p_o$$
$$+ U_{o1}\chi_1 U_{12}\chi_2 U_{21}\chi_1 U_{1s}p_o + U_{o2}\chi_2 U_{21}\chi_1 U_{12}\chi_2 U_{2s}p_o + \cdots. \qquad (12)$$

Continuing now with the example depicted in Fig. 2, if we place the same, or perhaps a different source dipole, say, $\boldsymbol{p}_o'$, at the observation point $\boldsymbol{r}_o$, and move the observer to $\boldsymbol{r}_s$, the contribution of each path through the system will be obtained by a similar procedure as before, except, of course, for the reversal of the direction in which each path is traversed. The contribution to the $E$-field by the reverse of the path taken in Eq.(11), for example, will be

$$\varepsilon_o E_o'(\boldsymbol{r}_s) = U_{s1}\chi_{E\_1} U_{12}\chi_{E\_2} U_{23}\chi_{E\_3} U_{34}\chi_{E\_4} U_{4o} p_o'. \qquad (13)$$

Note that the coefficient matrix of Eq.(13) is obtained from that in Eq.(11) by a simple transposition. The same procedure will apply to all the terms in the sum over the various paths. Therefore, the coefficient matrix relating the $E$-field observed at $\boldsymbol{r}_o$ when the source dipole is at $\boldsymbol{r}_s$, is the transpose of the coefficient matrix relating the $E$-field observed at $\boldsymbol{r}_s$ when the source dipole is at $\boldsymbol{r}_o$. This completes the proof of reciprocity in the case of electrically polarizable media excited by an electric dipole.

Given a source dipole $\boldsymbol{p}(t)$ aligned with the $x$-, $y$-, and $z$-axes in three separate experiments, one can make nine measurements of the various $E$-field components to obtain the complete 3×3 transfer matrix for the $E$-field produced by an electric dipole associated with given source and observer locations. (In a typical experiment, one would first orient $\boldsymbol{p}$ along the $x$-axis and measure $E_x$, $E_y$, $E_z$ at the observation point, then repeat the experiment with $\boldsymbol{p}$ aligned with the $y$-axis, and then again with $\boldsymbol{p}$ aligned with the $z$-axis.) According to the reciprocity theorem, the transpose of the above 3×3 matrix will be the transfer matrix when the source dipole (electric) is placed at the original observation point, and the original source location is chosen as the place to monitor the $E$-field.

Next, let the source dipole be a magnetic point-dipole, $\boldsymbol{m}_o$, while the surrounding material media are the same as before (i.e., containing electric dipoles with symmetric, time-independent susceptibility tensors). This time the observed field at $\boldsymbol{r}_o$ is the magnetic field $\mu_o \boldsymbol{H}$. The field arriving directly from the source is given by $\mu_o H_o = U_{os}m_o$, which remains intact when the source and observer positions are exchanged. For all the other paths that go through the surrounding media, the procedure is the same as before, except that Eqs.(11) and (13) will now become:

$$\mu_o H_o(\boldsymbol{r}_o) = -V_{o4}\chi_{E\_4} U_{43}\chi_{E\_3} U_{32}\chi_{E\_2} U_{21}\chi_{E\_1} V_{1s} m_o, \qquad (14a)$$

$$\mu_o H_o'(\boldsymbol{r}_s) = -V_{s1}\chi_{E\_1} U_{12}\chi_{E\_2} U_{23}\chi_{E\_3} U_{34}\chi_{E\_4} V_{4o} m_o'. \qquad (14b)$$

Note that the constant coefficients $Z_o$ and $-Z_o^{-1}$ which, according to Eqs.(9b) and (10b), must multiply $V_{1s}$ and $V_{o4}$, end up cancelling each other out, leaving only an inconsequential minus sign in front of both Eqs.(14a) and (14b). As before, the coefficient matrix of Eq.(14b) is the transpose of that in Eq.(14a). The observed $H$-field being the superposition of an infinite number of terms similar to those in Eqs.(14), we conclude that reciprocity continues to hold when the source dipole is magnetic and the observed field is the $H$-field.

By a completely analogous argument one can readily prove the reciprocity theorem in the case where the source dipole is electric while the observed field is magnetic, in which case the reverse path must have a magnetic dipole as the source and an $E$-field monitor for an observer.

**5. Reciprocity in systems containing both electric and magnetic media**. Let us first consider a stationary system of magnetic dipoles only, each dipole having its own symmetric, time-independent susceptibility tensor $\chi_{M\_n}$. Suppose these dipoles constitute the magnetic media that surround the source dipole at $r_s$ and the observation point $r_o$. The proof of reciprocity for this system follows exactly the same line of reasoning as was pursued in the preceding section in connection with media comprised solely of electric dipoles. For example, if the source is a magnetic dipole $m_o$ located at $r_s$, and we choose to monitor the $H$-field at $r_o$, all the matrices used along the various paths through the system will be $U$ matrices. If, however, the source is an electric point-dipole $p_o$, and we monitor the $E$-field at $r_o$, then the first and the last matrices associated with any path through the surrounding media will be $V$ matrices, while the matrices associated with interactions between pairs of magnetic dipoles will continue to be $U$ matrices.

In general, the surrounding media will contain both electric and magnetic dipoles, with symmetric, time-independent susceptibility tensors $\chi_{E\_n}$ and $\chi_{M\_m}$, respectively. The string of matrices associated with an arbitrary path from the source to the observer will then contain a $U$ matrix whenever a dipole is followed by the same type of dipole (i.e., electric dipole followed by electric dipole, or magnetic dipole followed by magnetic dipole), and a $V$ matrix whenever a magnetic dipole follows an electric dipole, or vice-versa. When the source and the observed field are of the same type, the $V$ matrices always occur in pairs, irrespective of the path taken from $r_s$ to $r_o$. To see this, consider a situation where the source is an electric dipole and the observer monitors the $E$-field. In this system, whenever a transition occurs from an electric dipole to a magnetic dipole along a given path, either the opposite transition takes place before the path terminates, or the last dipole to transmit its $E$-field to the observer will be a magnetic dipole. Either way, for every $V$ matrix that produces an $H$-field from an electric dipole, there will be a compensating $V$ matrix that generates an $E$-field from a magnetic dipole. Thus the coefficients $Z_o$ and $Z_o^{-1}$ do not fail to alternate along the path and always end up cancelling out.

It may happen that both an electric dipole and a magnetic dipole reside at a given point $r_n$ within the surrounding media. These dipoles, therefore, cannot influence each other directly, although they definitely affect each other through their interactions with other dipoles. Any path that includes $r_n$ must therefore choose between the electric dipole and the magnetic dipole residing at that location. If, before arriving at the observation point, the same path happens to visit $r_n$ again, the new choice between the two resident dipoles will be independent of the previous choice(s) made at that same location. In this way, there will be many paths through the system which are geometrically identical (because they contain the same sequence of points), yet they are physically distinct because of the choice(s) made with regard to selecting the electric or the magnetic dipole at those locations where both dipoles reside. In general, if the path contains $k$ locations where electric and magnetic dipoles overlap, there will be $2^k$ distinct physical sequences of dipoles corresponding to the same geometrical path. These are minor details that one must take into consideration when constructing a sequence of dipoles for a given path, but they do not alter the proof of reciprocity given in the preceding section.

Perhaps the easiest way to analyze a path through a system that contains both electric and magnetic dipoles is to keep track of *both* the $E$-field and the $H$-field along the entire path. The procedure is best described by a simple example. Suppose that, aside from $r_s$ and $r_o$, there are only two points along a path, $r_1$ and $r_2$, each containing an electric as well as a magnetic dipole. Forming 6×1 matrices such as $[p_n, m_n]$ and $[\varepsilon_o E_{on}, \mu_o H_{on}]$ to keep track of the fields and the dipoles, one can traverse the path $r_s \rightarrow r_1 \rightarrow r_2 \rightarrow r_o$ using the following sequence of operations:

$$\begin{pmatrix} \varepsilon_o E_o \\ \mu_o H_o \end{pmatrix}_O = \begin{pmatrix} U_{o2} & -Z_o^{-1}V_{o2} \\ Z_o V_{o2} & U_{o2} \end{pmatrix} \begin{pmatrix} \chi_{E\text{-}2} & 0 \\ 0 & \chi_{M\text{-}2} \end{pmatrix} \begin{pmatrix} U_{21} & -Z_o^{-1}V_{21} \\ Z_o V_{21} & U_{21} \end{pmatrix} \begin{pmatrix} \chi_{E\text{-}1} & 0 \\ 0 & \chi_{M\text{-}1} \end{pmatrix} \begin{pmatrix} U_{1s} & -Z_o^{-1}V_{1s} \\ Z_o V_{1s} & U_{1s} \end{pmatrix} \begin{pmatrix} p_o \\ m_o \end{pmatrix}_S$$

(15)

Carrying out the multiplications in the above equation, we find (for the path under consideration) all allowed sequences of the $U$ and $V$ matrices for different combinations of source dipoles and observed fields, namely,

$$\varepsilon_o E_o = (U_{o2}\chi_{E\text{-}2}U_{21}\chi_{E\text{-}1}U_{1s} - U_{o2}\chi_{E\text{-}2}V_{21}\chi_{M\text{-}1}V_{1s} - V_{o2}\chi_{M\text{-}2}V_{21}\chi_{E\text{-}1}U_{1s} - V_{o2}\chi_{M\text{-}2}U_{21}\chi_{M\text{-}1}V_{1s})p_o, \quad (16a)$$

$$\mu_o H_o = (U_{o2}\chi_{M\text{-}2}U_{21}\chi_{M\text{-}1}U_{1s} - V_{o2}\chi_{E\text{-}2}U_{21}\chi_{E\text{-}1}V_{1s} - V_{o2}\chi_{E\text{-}2}V_{21}\chi_{M\text{-}1}U_{1s} - U_{o2}\chi_{M\text{-}2}V_{21}\chi_{E\text{-}1}V_{1s})m_o, \quad (16b)$$

$$\mu_o H_o = Z_o(V_{o2}\chi_{E\text{-}2}U_{21}\chi_{E\text{-}1}U_{1s} + U_{o2}\chi_{M\text{-}2}V_{21}\chi_{E\text{-}1}U_{1s} + U_{o2}\chi_{M\text{-}2}U_{21}\chi_{M\text{-}1}V_{1s} - V_{o2}\chi_{E\text{-}2}V_{21}\chi_{M\text{-}1}V_{1s})p_o, \quad (16c)$$

$$\varepsilon_o E_o = -Z_o^{-1}(U_{o2}\chi_{E\text{-}2}U_{21}\chi_{E\text{-}1}V_{1s} + U_{o2}\chi_{E\text{-}2}V_{21}\chi_{M\text{-}1}U_{1s} + V_{o2}\chi_{M\text{-}2}U_{21}\chi_{M\text{-}1}U_{1s} - V_{o2}\chi_{M\text{-}2}V_{21}\chi_{E\text{-}1}V_{1s})m_o. \quad (16d)$$

In the above equations, each term corresponds to a different sequence of electric and magnetic dipoles along the path. The method automatically finds all possible combinations. Also note in Eqs. (16c) and (16d) that reciprocity holds even when the source is an electric dipole while the observed field at $r_o$ is magnetic, and vice-versa. The trick in such cases, when exchanging the positions of the source and the observer, is to switch the type of the source dipole ($p_o \leftrightarrow m_o$) and also the type of field that is being monitored ($E_o \leftrightarrow H_o$).

**6. Reciprocity in the presence of spatial dispersion**. In a medium exhibiting spatial dispersion, the dipoles respond not just to the local fields but also to fields at other locations within the material media. For example, the electric dipole residing at $r_n$ will respond to the local $E$-field with a susceptibility $\chi_{E\_nn}$, but also to the $E$-field at $r_m$ with a susceptibility $\chi_{E\_nm}$. In what follows, we will show that reciprocity will apply to such systems provided that $\chi_{nm} = \chi_{mn}^T$.

The proof of reciprocity in the presence of spatial dispersion follows the same line of reasoning as was used in the preceding sections. The main difference here is that, as we proceed along a given path from the source at $r_s$ to the observation point $r_o$, the contribution of each point such as $r_n$ to its immediate neighbor $r_{n+1}$ will be via the field at that point, say, $E(r_n)$, as well as radiation by the induced dipole at that point, $p(r_n)$. It is thus necessary at each step along the path to keep track of both $E(r_n)$ and the radiated field produced by $p(r_n)$ that reaches $r_{n+1}$.

A simple example will clarify the procedure. Suppose the chosen path is $r_s \to r_1 \to r_2 \to r_3 \to r_o$, and that the source dipole as well as the media dipoles are all electric. The source dipole $p_o$ excites the first dipole $p_1$ directly, which responds with a susceptibility $\chi_{11}$ to this excitation field. The effect of $p_1$ on $p_2$, however, will be via $E(r_1)$ as well as the radiated $E$-field of $p(r_1)$, that is, $p(r_2) = \varepsilon_o \chi_{21} E(r_1) + \chi_{22} U_{12} p(r_1)$. Similarly, the effect of $p_2$ on $p_3$ is via $E(r_2)$ as well as the radiated $E$-field of $p(r_2)$, that is, $p(r_3) = \varepsilon_o \chi_{32} E(r_2) + \chi_{33} U_{23} p(r_2)$. The dipole at $r_3$ then contributes $\varepsilon_o E(r_o) = U_{o3} p(r_3)$ to the field at the observation point. Using the 6×1 matrix $[\varepsilon_o E(r_n), p(r_n)]$ as a vehicle to keep track of both the local field and the local dipole, the observed field at the end of the $r_s \to r_1 \to r_2 \to r_3 \to r_o$ path may be written as follows:

$$\begin{pmatrix} \varepsilon_o E_o(r_o) \\ 0 \end{pmatrix}_o = \begin{pmatrix} 0 & U_{o3} \\ 0 & 0 \end{pmatrix} \begin{pmatrix} 0 & U_{32} \\ \chi_{32} & \chi_{33}U_{32} \end{pmatrix} \begin{pmatrix} 0 & U_{21} \\ \chi_{21} & \chi_{22}U_{21} \end{pmatrix} \begin{pmatrix} 0 & U_{1s} \\ 0 & \chi_{11}U_{1s} \end{pmatrix} \begin{pmatrix} 0 \\ p_o(r_s) \end{pmatrix}_S. \quad (17)$$

Straightforward multiplication of the above matrices yields

$$\varepsilon_o E_o(r_o) = U_{o3}(\chi_{32}U_{21}\chi_{11} + \chi_{33}U_{32}\chi_{21} + \chi_{33}U_{32}\chi_{22}U_{21}\chi_{11})U_{1s}p_o(r_s). \tag{18}$$

The first term on the right-hand-side of Eq.(18) represents the contribution of the dipole at $r_3$ induced by the $E$-field at $r_2$, which is the result of radiation from the dipole at $r_1$, which dipole is directly excited by the source $p_o$. The second term represents the contribution to the observed $E$-field by the dipole at $r_3$ induced by the local field, which is produced by radiation from the dipole at $r_2$, which is in turn excited by the $E$-field that reaches $r_1$ directly from the source $p_o$. The last term is the contribution to the observed $E$-field by the dipole at $r_3$ induced by the local field, which has arrived there from the dipole at $r_2$, also induced by the local field, which is the result of propagation from the dipole at $r_1$, also induced by the local field, which field has propagated to $r_1$ from the source $p_o$. It is easy to see from Eq.(18) that reciprocity will hold if, upon reversing the propagation path (i.e., going from $r_o$ to $r_s$), $\chi_{32}^T$ would become $\chi_{23}$ and $\chi_{21}^T$ would become $\chi_{12}$. Needless to say, the local susceptibility tensors, $\chi_{11}$, $\chi_{22}$, and $\chi_{33}$ must also be symmetric, as before. In general, therefore, the condition for reciprocity in the presence of spatial dispersion is $\chi_{E\_mn} = \chi_{E\_nm}^T$.

The above result can readily be generalized to the case of material media containing both electric and magnetic dipoles. With the obvious exception of the source location $r_s$ and the observation point $r_o$, every point along any path through the media must now be host to both an electric dipole and a magnetic dipole. As discussed at the end of Section 5, at each point $r_n$ along the geometric path, for any particular realization of a string of dipoles, we must choose the resident electric dipole or the resident magnetic dipole (but not both). If the path contains $k$ such points, there will be $2^k$ physically distinct paths corresponding to the same geometrical path. Subsequently, in passing from $r_n$ to $r_{n+1}$ along any such string of dipoles, we must allow the $E(r_n)$ to excite the next dipole only if the next dipole is an electric dipole. Similarly, $H(r_n)$ can excite the dipole at $r_{n+1}$ only if that dipole is magnetic. However, the radiation from either type of dipole at $r_n$ contains both an $E$-field and an $H$-field, one of which will excite the dipole at $r_{n+1}$, depending on whether the dipole residing at $r_{n+1}$ is electric or magnetic. A good way to keep track of all the interactions is via the $12 \times 1$ matrices $[\varepsilon_o E(r_n), \mu_o H(r_n), p(r_n), m(r_n)]$ and the corresponding $12 \times 12$ matrices composed of $U_{mn}$, $V_{mn}$, and the relevant susceptibility tensors that propagate the fields and the dipole strengths from $r_n$ to $r_m$. In this way, any spatially dispersive system containing both electric and magnetic dipoles can be properly modeled with the aid of our discrete dipoles. Following the same line of reasoning as in the earlier discussions, one can now show that reciprocity will hold provided that $\chi_{E\_mn} = \chi_{E\_nm}^T$ and $\chi_{M\_mn} = \chi_{M\_nm}^T$.

**7. Comparison with the standard proof of reciprocity**. The original proof of reciprocity for linear media in the absence of spatial dispersion is based on Maxwell's macroscopic equations through the following line of argument. Suppose the only source of radiation in the system is an oscillating electric point-dipole of magnitude $p_o$ and frequency $\omega_o$, located at $r_s$. The polarization density distribution, $P_1(r,t)$, may then be split into a source term, $p_o\delta(r-r_s)\exp(-i\omega_o t)$, and a media term, $\varepsilon_o \chi_E(r)E_1(r)\exp(-i\omega_o t)$, where the susceptibility tensor $\chi_E(r)$ is symmetric at all points $r$ within the surrounding media. In this first arrangement involving source at $r_s$ and observation point at $r_o$, the $E$-field magnitude is denoted by $E_1(r)$. Also assumed is a magnetization distribution throughout the surrounding media given by $\mu_o \chi_M(r)H_1(r)\exp(-i\omega_o t)$. Maxwell's curl equations may now be written as follows:

$$\nabla \times H_1(r) = -i\omega_o p_o \delta(r-r_s) - i\omega_o \varepsilon_o [1+\chi_E(r)] E_1(r), \tag{19a}$$

$$\nabla \times E_1(r) = i\omega_o \mu_o [1+\chi_M(r)] H_1(r). \tag{19b}$$

In the reverse configuration, let the source be $p_o' \delta(r-r_o)\exp(-i\omega_o t)$, which is a point-dipole of magnitude $p_o'$ located at $r_o$. The $E$- and $H$-fields in this case will be identified by the subscript 2, and the curl equations will be

$$\nabla \times H_2(r) = -i\omega_o p_o' \delta(r-r_o) - i\omega_o \varepsilon_o [1+\chi_E(r)] E_2(r), \tag{20a}$$

$$\nabla \times E_2(r) = i\omega_o \mu_o [1+\chi_M(r)] H_2(r). \tag{20b}$$

We now multiply $E_2(r)$ into Eq.(19a) and $H_1(r)$ into Eq.(20b), then subtract the latter equation from the former, to obtain

$$E_2(r) \cdot [\nabla \times H_1(r)] - H_1(r) \cdot [\nabla \times E_2(r)] = \nabla \cdot [H_1(r) \times E_2(r)]$$
$$= -i\omega_o E_2(r) \cdot p_o \delta(r-r_s) - i\omega_o \varepsilon_o E_2(r)[1+\chi_E(r)]E_1(r) - i\omega_o \mu_o H_1(r)[1+\chi_M(r)]H_2(r). \tag{21a}$$

In like manner, we multiply $H_2(r)$ into Eq.(19b) and $E_1(r)$ into Eq.(20a), then subtract the former equation from the latter, to obtain

$$E_1(r) \cdot [\nabla \times H_2(r)] - H_2(r) \cdot [\nabla \times E_1(r)] = \nabla \cdot [H_2(r) \times E_1(r)]$$
$$= -i\omega_o E_1(r) \cdot p_o' \delta(r-r_o) - i\omega_o \varepsilon_o E_1(r)[1+\chi_E(r)]E_2(r) - i\omega_o \mu_o H_2(r)[1+\chi_M(r)]H_1(r). \tag{21b}$$

If Eq.(21b) is now subtracted from Eq.(21a), the symmetry of $\chi_E(r)$ and $\chi_M(r)$ causes the terms containing these tensors to cancel out. The resulting equation will be

$$\nabla \cdot [E_1(r) \times H_2(r) - E_2(r) \times H_1(r)] = i\omega_o [E_1(r) \cdot p_o' \delta(r-r_o) - E_2(r) \cdot p_o \delta(r-r_s)]. \tag{22}$$

Upon integrating Eq.(22) over the entire space, the integrated divergence on the left-hand side, in accordance with Gauss's theorem, reduces to the surface integral of $E_1 \times H_2 - E_2 \times H_1$ at infinity, which can be argued to approach zero when the integration surface is sufficiently far from all sources of radiation (this involves a non-trivial argument). Setting the integral of the right-hand side of Eq.(22) equal to zero then yields $E_1(r_o) \cdot p_o' = E_2(r_s) \cdot p_o$, which is the statement of reciprocity for the case under consideration.

In a completely analogous way, we let the source at $r_s$ remain the electric point-dipole $p_o$, but, in the reverse path, we substitute a magnetic point-dipole $m_o'$ for the source at $r_o$. Equations (19) thus remain the same, but, in Eqs.(20), the source term moves to the second curl equation and appears as $+i\omega_o m_o' \delta(r-r_o)$. The rest of the proof remains unchanged, and the final result will be $H_1(r_o) \cdot m_o' = -E_2(r_s) \cdot p_o$. Similarly, when both sources are magnetic dipoles, the above derivation yields $H_1(r_o) \cdot m_o' = H_2(r_s) \cdot m_o$.

These classical results, of course, are identical with those obtained using the proposed scheme of the present paper. One advantage of the proposed method is that it can be readily generalized to the case of spatially dispersive media, as was shown in Section 6. Another advantage is that it does *not* require the surface integral of $E_1 \times H_2 - E_2 \times H_1$ to vanish in the limit when the integration surface moves to infinity.

**8. Comparison with the proof of reciprocity based on Green's functions**. A powerful method of proving reciprocity involves the use of Green's functions. In the following paragraphs we derive the discrete version of Green's function for a simple surrounding medium that contains only two pairs of point-dipoles, ($p_1,m_1$) co-located at $r_1$, and ($p_2,m_2$) co-located at $r_2$. Despite its simplicity, this example embodies the essence of the Green's function approach to solving electromagnetic problems.

Let the pair of point-dipoles ($p_o,m_o$) be co-located at the source point $r_s$, oscillating with the constant frequency $\omega_o$. The electromagnetic radiation from these source dipoles will excite the dipoles at $r_1$ and $r_2$, which have local susceptibilities ($\chi_{E\_11}, \chi_{M\_11}$) and ($\chi_{E\_22}, \chi_{M\_22}$), respectively. Additionally, to introduce spatial dispersion into this discrete system, we define the susceptibilities ($\chi_{E\_12}, \chi_{M\_12}$) for the dipoles at $r_1$ responding to the $E$- and $H$-fields at $r_2$, and also ($\chi_{E\_21}, \chi_{M\_21}$) for the dipoles at $r_2$ responding to the $E$- and $H$-fields at $r_1$. To keep track of the fields as well as the dipole strengths at various locations, we use the $1 \times 12$ vector [$\varepsilon_o E(r), \mu_o H(r), p(r), m(r)$], writing the fields and the induced dipoles at $r_1$ as follows:

$$\begin{pmatrix} \varepsilon_o E_1 \\ \mu_o H_1 \\ p_1 \\ m_1 \end{pmatrix} = \begin{pmatrix} 0 & 0 & U_{12} & -Z_o^{-1} V_{12} \\ 0 & 0 & Z_o V_{12} & U_{12} \\ \chi_{E-12} & 0 & \chi_{E\_11} U_{12} & -Z_o^{-1} \chi_{E\_11} V_{12} \\ 0 & \chi_{M-12} & Z_o \chi_{M\_11} V_{12} & \chi_{M\_11} U_{12} \end{pmatrix} \begin{pmatrix} \varepsilon_o E_2 \\ \mu_o H_2 \\ p_2 \\ m_2 \end{pmatrix} + \begin{pmatrix} U_{1S} \\ Z_o V_{1S} \\ \chi_{E\_11} U_{1S} \\ Z_o \chi_{M\_11} V_{1S} \end{pmatrix} p_o + \begin{pmatrix} -Z_o^{-1} V_{1S} \\ U_{1S} \\ -Z_o^{-1} \chi_{E\_11} V_{1S} \\ \chi_{M\_11} U_{1S} \end{pmatrix} m_o$$

(23)

Introducing an obvious notation to express the above equation in a more compact form, Eq. (23) may be rewritten

$$C_1 = W_{12} C_2 + X_{1S} p_o + Y_{1S} m_o. \tag{24a}$$

In like manner, the corresponding equation for the fields and dipoles at $r_2$, excited by those at $r_1$ and $r_s$, will be

$$C_2 = W_{21} C_1 + X_{2S} p_o + Y_{2S} m_o. \tag{24b}$$

The solution to these equations is readily obtained by solving the matrix equations for [$C_1, C_2$], namely,

$$\begin{pmatrix} C_1 \\ C_2 \end{pmatrix} = \begin{pmatrix} I & -W_{12} \\ -W_{21} & I \end{pmatrix}^{-1} \begin{pmatrix} X_{1S} \\ X_{2S} \end{pmatrix} p_o + \begin{pmatrix} I & -W_{12} \\ -W_{21} & I \end{pmatrix}^{-1} \begin{pmatrix} Y_{1S} \\ Y_{2S} \end{pmatrix} m_o. \tag{25}$$

The fields produced at the observation point $r_o$ by the dipoles at $r_1$ and $r_2$ may now be obtained by multiplying into [$C_1, C_2$] the following $6 \times 24$ propagation matrix:

$$\begin{pmatrix} 0 & 0 & U_{o1} & -Z_o^{-1} V_{o1} & 0 & 0 & U_{o2} & -Z_o^{-1} V_{o2} \\ 0 & 0 & Z_o V_{o1} & U_{o1} & 0 & 0 & Z_o V_{o2} & U_{o2} \end{pmatrix}. \tag{26}$$

In this way we find the fields at the observation point due to the excitation of the media by the source dipoles. We must also remember to add to these fields the direct contribution from the source dipoles ($p_o,m_o$) located at $r_s$.

The continuum analog of the inverted square matrix appearing in Eq.(25) is a Green's function. Given the symmetries inherent in this matrix as well as in the propagation matrices between the source and the media, and also those between the media and the observation point, it is possible to prove reciprocity in a system with symmetric susceptibility tensors, namely, $\chi_{E\_mn} = \chi_{E\_nm}^T$ and $\chi_{M\_mn} = \chi_{M\_nm}^T$ [10,11]. The proof is considerably more complicated, however, than the one presented here in Section 6. The advantage of our method of proof is that it only requires "partial" fields (and the correspondingly induced partial dipoles) at each point within the surrounding media. In contrast, Eq.(25) contains the complete solution $[C_1, C_2]$ of Maxwell's equations for all the dipoles of the surrounding media. In other words, there is more complexity in the Green's function approach to proving reciprocity than is actually needed for the theorem.

**9. Summary and concluding remarks**. In this paper we have introduced an elementary yet powerful approach to proving the reciprocity theorem of classical electrodynamics, as well as gaining a better understanding of the physical basis of the theorem. Our results pertain to the case when an electric or a magnetic point-dipole, $\boldsymbol{p}_o \exp(-i\omega_o t)$ or $\boldsymbol{m}_o \exp(-i\omega_o t)$, located at a source point $\boldsymbol{r}_s$ and surrounded by linear, time-invariant media specified by their electric and magnetic susceptibilities, $\varepsilon_o \chi_E(\boldsymbol{r}, \boldsymbol{r}')$ and $\mu_o \chi_M(\boldsymbol{r}, \boldsymbol{r}')$, produces the EM fields $\boldsymbol{E}_o(\boldsymbol{r}_o) \exp(-i\omega_o t)$ and $\boldsymbol{H}_o(\boldsymbol{r}_o) \exp(-i\omega_o t)$ at an observation point $\boldsymbol{r}_o$. If the source and the observer positions are exchanged, and if a different dipole, either $\boldsymbol{p}'_o \exp(-i\omega_o t)$ or $\boldsymbol{m}'_o \exp(-i\omega_o t)$, is placed at $\boldsymbol{r}_o$, the observed fields at $\boldsymbol{r}_s$ will be $\boldsymbol{E}'_o(\boldsymbol{r}_s) \exp(-i\omega_o t)$ and $\boldsymbol{H}'_o(\boldsymbol{r}_s) \exp(-i\omega_o t)$. According to the reciprocity theorem, when the susceptibility tensors are symmetric, that is, when $\chi_E(\boldsymbol{r}, \boldsymbol{r}') = \chi_E^T(\boldsymbol{r}', \boldsymbol{r})$ and $\chi_M(\boldsymbol{r}, \boldsymbol{r}') = \chi_M^T(\boldsymbol{r}', \boldsymbol{r})$, the following relations hold among the various sources and observed fields.

i) $\boldsymbol{p}'_o \cdot \boldsymbol{E}_o(\boldsymbol{r}_o) = \boldsymbol{p}_o \cdot \boldsymbol{E}'_o(\boldsymbol{r}_s)$ when the source dipoles in both forward and reverse paths are electric.

ii) $\boldsymbol{m}'_o \cdot \boldsymbol{H}_o(\boldsymbol{r}_o) = \boldsymbol{m}_o \cdot \boldsymbol{H}'_o(\boldsymbol{r}_s)$ when the source dipoles in both forward and reverse paths are magnetic.

iii) $\boldsymbol{m}'_o \cdot \boldsymbol{H}_o(\boldsymbol{r}_o) = \boldsymbol{p}_o \cdot \boldsymbol{E}'_o(\boldsymbol{r}_s)$ when the source dipole in the forward path is electric, while that in the reverse path is magnetic.

iv) $\boldsymbol{p}'_o \cdot \boldsymbol{E}_o(\boldsymbol{r}_o) = \boldsymbol{m}_o \cdot \boldsymbol{H}'_o(\boldsymbol{r}_s)$ when the source dipole in the forward path is magnetic, while that in the reverse path is electric.

It should be obvious that the above results can be readily generalized to the case of extended sources, as an extended source is nothing but a collection of dipoles at different spatial locations, whose observed field at any given point $\boldsymbol{r}_o$ is the linear superposition of the fields produced by each and every dipole associated with the source.

In the literature, the Rayleigh-Carson-Lorentz reciprocity theorem is often stated in terms of the source current density $\boldsymbol{J}_o(\boldsymbol{r}) \exp(-i\omega_o t)$ and the observed $E$-field $\boldsymbol{E}_o(\boldsymbol{r}) \exp(-i\omega_o t)$, as follows:

$$\int_{V_1} \boldsymbol{J}_o(\boldsymbol{r}) \cdot \boldsymbol{E}'_o(\boldsymbol{r}) \mathrm{d}\boldsymbol{r} = \int_{V_2} \boldsymbol{J}'_o(\boldsymbol{r}) \cdot \boldsymbol{E}_o(\boldsymbol{r}) \mathrm{d}\boldsymbol{r}. \tag{27}$$

In the above equation, the source in the forward path occupies a volume $V_1$, while that in the reverse path occupies a volume $V_2$. Equation (27) is equivalent to our version of the theorem, $\boldsymbol{p}_o \cdot \boldsymbol{E}'_o(\boldsymbol{r}_s) = \boldsymbol{p}'_o \cdot \boldsymbol{E}_o(\boldsymbol{r}_o)$, for an extended source, the reason being that, so far as Maxwell's equations are concerned, $\boldsymbol{J}_o(\boldsymbol{r}) \exp(-i\omega_o t)$ and $\partial \boldsymbol{P}(\boldsymbol{r}, t)/\partial t = -i\omega_o \boldsymbol{P}_o(\boldsymbol{r}) \exp(-i\omega_o t)$ are indistinguishable, provided, of course, that $\omega_o \neq 0$. This is also the reason why we have left the term $\boldsymbol{J}_{\text{free}}(\boldsymbol{r},t)$ out of the Maxwell-Ampere equation $\nabla \times \boldsymbol{H}(\boldsymbol{r},t) = \boldsymbol{J}_{\text{free}}(\boldsymbol{r},t) + \partial \boldsymbol{D}(\boldsymbol{r},t)/\partial t$; see Eq. (19a). Any part of $\boldsymbol{J}_{\text{free}}(\boldsymbol{r},t)$ that is associated with the source may be replaced with $\partial \boldsymbol{P}(\boldsymbol{r},t)/\partial t$, and any part of it that is associated with the surrounding (linear) media through the conductivity tensor $\sigma(\boldsymbol{r},\boldsymbol{r}')$ may be replaced with electric dipoles having susceptibility $\chi_E(\boldsymbol{r},\boldsymbol{r}') = i\sigma(\boldsymbol{r},\boldsymbol{r}')/(\varepsilon_o \omega_o)$. The statement of the theorem in terms of electric point-dipoles is, therefore, completely equivalent to that in terms of current density distributions, as in Eq. (27).

A version of the reciprocity theorem, known as the Feld-Tai lemma [8,9], is usually stated as follows:

$$\int_{V_1} \boldsymbol{J}_o(\boldsymbol{r}) \cdot \boldsymbol{H}'_o(\boldsymbol{r}) d\boldsymbol{r} = \int_{V_2} \boldsymbol{J}'_o(\boldsymbol{r}) \cdot \boldsymbol{H}_o(\boldsymbol{r}) d\boldsymbol{r}. \tag{28}$$

The Feld-Tai lemma is not as general as the Rayleigh-Carson-Lorentz theorem stated in Eq. (27). For example, in the proof provided by C. T. Tai [9], some of the surrounding media in the forward path must be replaced by "complementary" media in the reverse path. The electric and magnetic susceptibilities of the surrounding media must be scalar entities (i.e., isotropic media), the dielectric and magnetic media must be piece-wise homogeneous (e.g., stratified media), and, in going from the forward to the reverse path, the susceptibilities of these isotropic and stratified media must be modified in accordance with a certain algorithm. Moreover, any perfect electrical conductor in the forward path becomes a perfect magnetic conductor in the reverse path.

We have *not* been able to prove the Feld-Tai lemma of Eq. (28) using our proposed method. However, we can prove Eq. (28) under the far less stringent condition that, in going from the forward to the reverse path, all the surrounding media be replaced with their complements, in the sense that the electric and magnetic susceptibility tensors $\chi_E(\boldsymbol{r},\boldsymbol{r}')$ and $\chi_M(\boldsymbol{r},\boldsymbol{r}')$ be exchanged. This version of the reciprocity theorem can be proven in very much the same way as the Rayleigh-Carson-Lorentz version was proven in the preceding sections. Once again, the only constraint on susceptibility tensors is the requirement of symmetry, namely, $\chi_E(\boldsymbol{r},\boldsymbol{r}') = \chi_E^T(\boldsymbol{r}',\boldsymbol{r})$ and $\chi_M(\boldsymbol{r},\boldsymbol{r}') = \chi_M^T(\boldsymbol{r}',\boldsymbol{r})$.

Finally, it must be pointed out that, in recent years, the angular spectrum of the electromagnetic field emitted by dipoles has been used to ascertain for all points in space (including the near field) the reciprocity and unitarity of the scattering (or *S*) matrix [19-21]. This important extension of the reciprocity theorem confirms the conservation of "information" in the near field, where evanescent and inhomogeneous fields predominate. The extension has also shed light on the connections among reciprocity, unitarity, and time-reversal invariance in classical optics. Our proposed formulation of reciprocity in the present paper is in complete accord with the aforementioned extension of the classical theorem to situations involving the near-field. This should be evident from Eqs. (9) and (10), which are the exact solutions of Maxwell's equations for radiating point-dipoles; solutions that are applicable to *all* points in the surrounding space, from the immediate vicinity of the dipole all the way across to the far field.

**Acknowledgement**. The authors are grateful to Pui-Tak Leung for many helpful discussions. We also thank the anonymous referee who drew our attention to references 19-21. One of the authors (M.M.) also would like to acknowledge the support from the National Science Council of Taiwan while he was on sabbatical leave at the National Taiwan University in Taipei.